# Electronic correlation effects in the $Cr_2GeC$ $M_{n+1}AX_n$ phase

Maurizio Mattesini[1] and Martin Magnuson[2]

[1]*Departamento de Física de la Tierra, Astronomía y Astrofísica I, Universidad Complutense de Madrid, E-28040 Madrid, Spain. Instituto de Geociencias (UCM-CSIC), Facultad de Ciencias Físicas, Plaza de Ciencias 1, 28040-Madrid, Spain.*

[2]*Department of Physics, Chemistry and Biology (IFM), Linköping University, SE-58183 Linköping, Sweden.*

*Email: mmattesi@fis.ucm.es, m.mattesini@igeo.ucm-csic.es*

## Abstract

The magnetic properties, electronic band structure and Fermi surfaces of the hexagonal $Cr_2GeC$ system have been studied by means of both generalized gradient approximation (GGA) and the +U corrected method (GGA+U). The effective *U* value has been computed within the augmented plane-wave theoretical scheme by following the constrained density functional theory formalism of Anisimov *et al.* [1991 *Phys. Rev. B* 45, 7570]. On the basis of our GGA+U calculations, a compensated anti-ferromagnetic spin ordering of Cr atoms has been found to be the ground state solution for this material, where a Ge-mediated super-exchange coupling is responsible for an opposite spin distribution between the ABA stacked in-plane Cr-C networks. Structural properties have also been tested and found to be in good agreement with the available experimental data. Topological analysis of Fermi surfaces have been used to qualitatively address the electronic transport properties of $Cr_2GeC$ and found an important asymmetrical carrier-type distribution within the hexagonal crystal lattice. We conclude that an appropriate description of the strongly correlated Cr-*d* electrons is an essential issue for interpreting the material properties of this unusual Cr-based *MAX*-phase.

## 1. Introduction

The $M_{n+1}AX_n$ or *MAX*-phases are layered hexagonal solids with unusual and sometimes unique combination of properties [2]. They are made of an early transition metal *M*, an *A*-group element (group *III*, *IV*, *V*, or *VI* element), and by an *X* element that is either C or N. They have attracted much attention due to the peculiar combination of properties that are normally associated with either metals or ceramics. Just like metals, they are readily machinable, electrically and thermally conductive, not susceptible to thermal shock, plastic at high temperature and exceptionally damage tolerant. On the other hand, they are elastically rigid, lightweight, creep and fatigue resistant as ceramic materials. It is therefore not surprising that there has been a rapid increase of research activities on *MAX* phases by both experimental and theoretical works during recent years [2, 4-6]. A magnetic *MAX* phase that could potentially give rise to functional materials for spintronics applications [3] has also long been searched. However, to our knowledge, despite many attempts and efforts, none of the synthesized phases have been found to possess stable magnetic features.

Among the known $M_{n+1}AX_n$ phases, $Cr_2GeC$ (Fig. 1) is a relatively little studied member. It has the highest thermal expansion coefficient among all the present known *MAX* phases [7-9], high resistivity, a positive Seebeck coefficient both in- and out-of-plane [10], and a negative Hall coefficient. The calculated electronic density of states (DOS) at the Fermi level ($E_F$) is considerably underestimated [11] and there is a large and anisotropic electron-phonon coupling [12]. In general, Cr-containing $M_{n+1}AX_n$-phases have unusually large DOS at the stemming from the electronic *d*-states of the transition metal. In fact, for $Cr_2GeC$ the DOS at is by far the highest (22 $eV^{-1}$ $cell^{-1}$) measured among the -phases and for $Cr_2AlC$ (14.6 $eV^{-1}$ $cell^{-1}$) it is the second highest. The carrier mobility is, however, rather limited in Cr-based *MAX* phases compared to Ti-containing ones, due to their strongly localized Cr *d*-states. The significantly correlated nature of the Cr *d*-electrons also make the magnetic coupling and ferromagnetic/anti-ferromagnetic ordering largely unknown and rather complicated to establish [8]. Moreover, there is some disagreement in the literature about the experimentally determined values for both bulk modulus and equilibrium volume [13,14].

For all these reasons, a comprehensive theoretical study is needed to correctly address the material properties of this unusual *MAX* phase. Specifically, the aim of this work is to study the effect of correlation on the electronic structure and material properties of the $Cr_2GeC$ *MAX*-phase. Particular attention has been given to the ground-state magnetic spin ordering, electro-structural correlations and transport properties. We observe that ferromagnetic Cr layers are anti-ferromagnetically coupled together via an interleaved Ge-atom, assembling a multilayer material that could, in principle, be tuned to provide thermodynamic stable magnetic *MAX*-phases.

## 2 Computational details

### 2.1 First-principles calculations

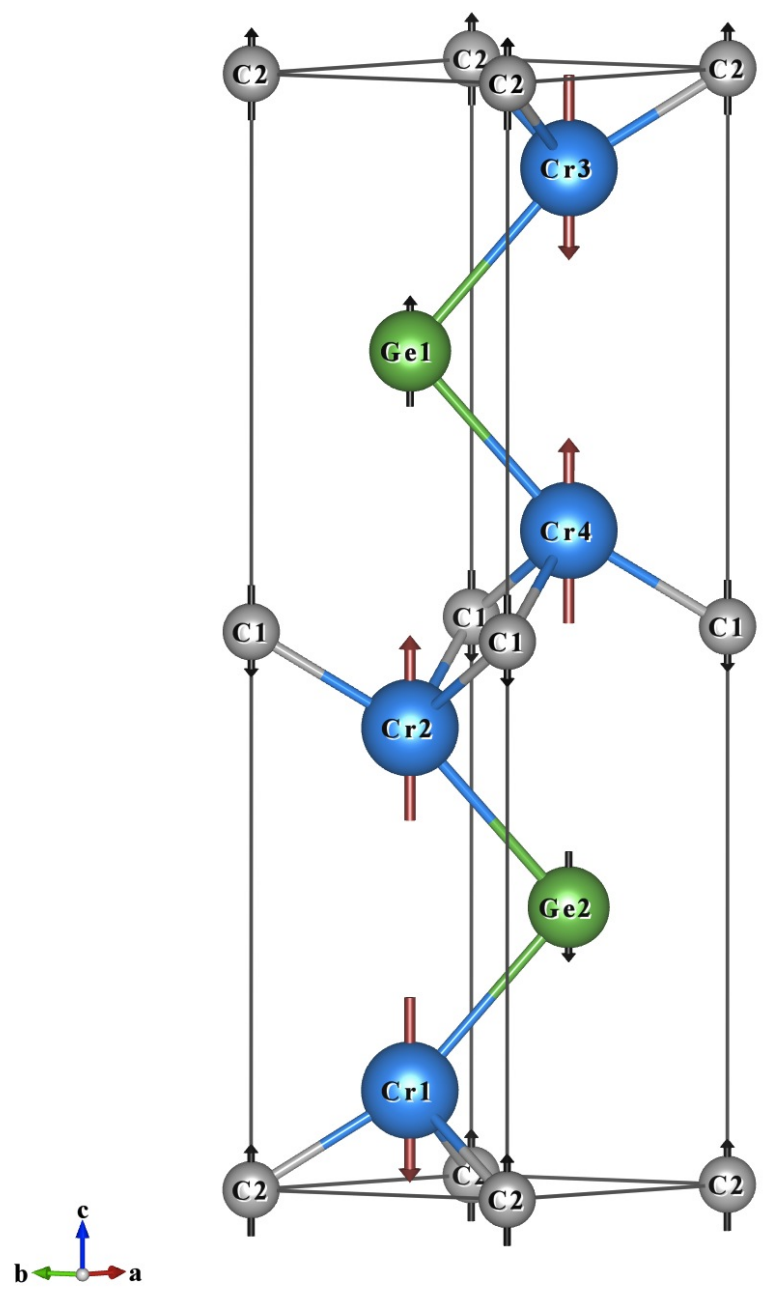

**Figure 1:** Atomic model of the hexagonal crystal structure of $Cr_2GeC$ with space group $P3m1$ (156) and 8-atoms per unit-cell. Chromium (Cr), germanium (Ge), and carbon (C) atoms are depicted in blue (large spheres), green (medium spheres), and gray (small spheres), respectively. The large red arrows drawn at the $Cr_1 \cdots Cr_4$ atoms indicate the ground-state magnetic spin configuration inside the unit cell. Small black arrows represent the fractional magnetic spin moments localized at the Ge and C sites. The VESTA visualization software [15] was used to generate the present figure.

The electronic structure of $Cr_2GeC$ was computed within the wien2k code [16] employing the density-functional [17,18] augmented plane wave plus local orbital (APW+lo) computational scheme. The APW+lo method expands the Kohn-Sham orbitals in atomic-like orbitals inside the muffin-tin (MT) atomic spheres and plane waves in the interstitial region. The Kohn-Sham equations were solved by means of the Wu-Cohen generalized gradient approximation (GGA-WC) [19,20] for the exchange-correlation (*xc*) potential. For a variety of materials it improves the equilibrium lattice constants and bulk moduli significantly over the local density approximation (LDA) [18], and performs pretty well for the $Cr_2GeC$ material (see results in Table 2). The latter is the main reason that motivated our choice to adopt the Wu-Cohen approximation in studying this Cr-based *MAX* phase.

A plane-wave expansion with $R_{MT} \cdot K_{max}=10$ was used in the interstitial region, while the potential and the charge density were Fourier expanded up to $G_{max}=12$. The modified tetrahedron method [21] was applied to integrate inside the Brillouin zone (*BZ*), and a **k**-point sampling with a 35×35×7 Monkhorst-Pack [22] mesh in the full BZ (corresponding to 786 irreducible **k**-points) was considered satisfactory for the hexagonal $Cr_2GeC$ system. Magnetic ground-state properties and electronic band structure features were studied using the relaxed unit cell parameters. All the spin-polarized calculations were charge converged up to $10^{-4}$ *e*.

Convergent and smooth Fermi surfaces (FSs) were achieved by sampling the whole BZ with 10000 **k**-points along the 35×35×7 Monkhorst-Pack grid. The presented FS plots were then generated with the help of the XCRYSDEN graphical user interface code [23] applying the tricubic spline interpolation with a degree of five.

## 2.2 Searching for an effective Hubbard *U*-value

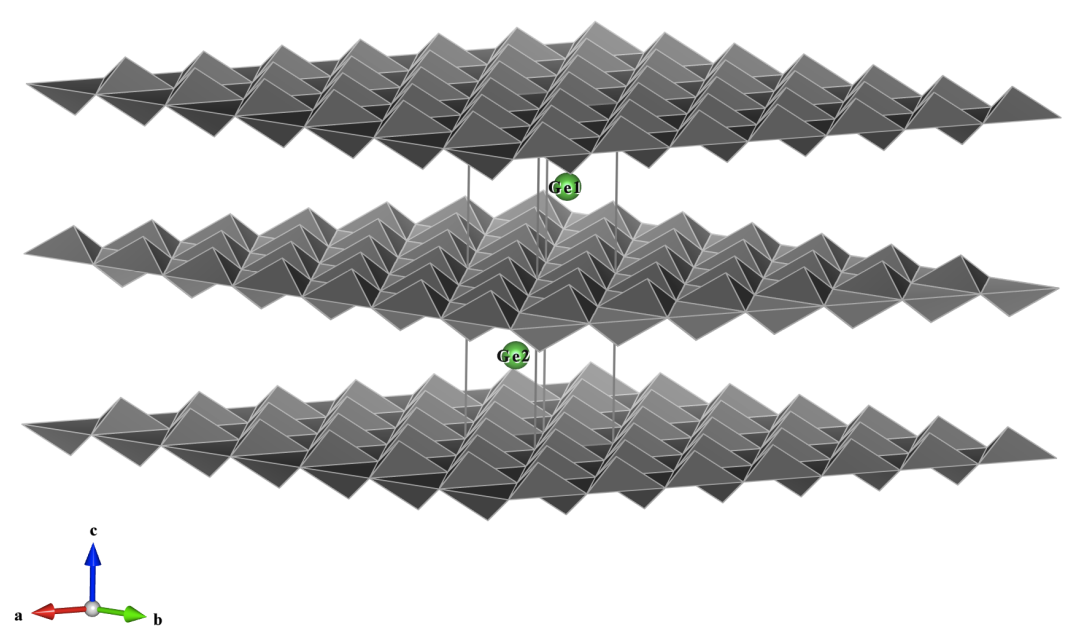

**Figure 2:** Polyhedral model for the $Cr_2GeC$ phase showing the Cr-C layers (gray) that are propagating along the *a*-*b* crystal plane and the interleaved Cr atoms (green). Each polyhedral skeleton consists of three C atoms at the base and a Cr atom at the vertex. Note the alternating (up/down) vertex distribution within each layer that produce an ABA stacking order of layers along the *c*-axis. The Cr atom sitting on the polyhedral vertices of the middle (bottom-top) Cr-C network is $Cr_4$ ($Cr_1$) for the upward polyhedra and $Cr_2$ ($Cr_3$) for its downward counterpart.

Density functional theory (DFT) is an upright method for computing ground-state properties of solids with feeble electronic correlations. However, this method fails to describe systems with intermediate and strong electron correlations, such as transition-metal oxides, Kondo systems and rare earths. Such a short-coming description is due to the spurious self-interaction. Therefore, these materials are very often investigated by means of a phenomenological many-body Hamiltonian such as the Hubbard model [24], where the effective on-site Coulomb interaction is an empirical parameter (*U*) that permit to reproduce the experimental results of interest. Hence, by using this approach, the correct determination of *U* represents a critical issue because many properties, such as magnetism, can vary in *MAX* phases with the value of *U*.

When employing the local density approximation for the *xc*-part, then the LDA+U method indicates that an orbital-dependent field has been introduced to correct for self-interaction [25]. Particularly, a set of atomic-like orbitals is treated with an orbital-dependent potential with an associated on-site Coulomb (*U*) and exchange (*J*) interactions. Since in LDA, the electron-electron interactions have already been considered in a mean-field way, one has to identify the parts that occur twice and apply a double-counting (DC) correction. To overcome such a problem, in the non-spherical part of potential we used $U_{eff} = U - J$ [26], setting therefore $J$=0. Although several different ways of correcting for DC are existing [25,1,27,28], we here use what has been referred to as the SIC method introduced by Anisimov *et al.* [25].

The physical meaning of the *U* parameter was defined by Anisimov and Gunnarsson [1], who described it as the Coulombic energy cost of placing two electrons on the same site. In an atom, *U* simply corresponds to the unscreened Slater-integrals, whereas in solids the is much smaller because of screening effects. The Hubbard *U* depends on the type of crystal structure, *d* electron number, *d* orbital filling and most generally on the degree of electronic localization.

Using the method of Anisimov and Gunnarsson (sometimes called *constrained DFT formalism*), the *U*-value has been calculated for the Cr atom in the hexagonal $Cr_2GeC$ structure. Two kinds of calculations were performed on a 2×2×1 supercell each with one impurity site forced to have the *d*-configuration as shown in eq. (1)

$$U_{\text{eff}} = \varepsilon_{3d\uparrow}\left(\frac{n+1}{2}, \frac{n}{2}\right) - \varepsilon_{3d\uparrow}\left(\frac{n+1}{2}, \frac{n}{2} - 1\right) - E_{\text{F}}\left(\frac{n+1}{2}, \frac{n}{2}\right) + E_{\text{F}}\left(\frac{n+1}{2}, \frac{n}{2} - 1\right) \quad (1)$$

where $\varepsilon_{3d\uparrow}$ is the spin-up 3*d* eigenvalue and E the Fermi energy. The *d*-character of the augmented plane waves at the chromium impurity sites was eliminated by setting the *d*-linearization energy far above the Fermi level [$E_{(l=2)}$=20.30 Ry]. Using eq. (1) we computed a $U_{eff}$ value of 2.09 eV ($t_{2g}$) and 2.33 eV ($e_g$) for the chosen $GGA^{WC}$ *xc*-functional. Table 1 shows the obtained Hubbard *U* parameter for various exchange-correlation potentials. In agreement with previous studies [29], the effective interaction between *d* electrons in $e_g$ orbitals is larger than that in $t_{2g}$ orbitals.

**Table 1:** Calculated on-site Coulomb value ( in eV) for different exchange-correlation functionals.

| | LDA | $GGA^{WC}$ | $GGA^{PBE}$ |
|---|---|---|---|
| $U_{\text{eff}}$ ($e_g$) | 2.14 | 2.33 | 2.08 |
| $U_{\text{eff}}$ ($t_{2g}$) | 2.04 | 2.09 | 2.03 |

# 3 Results

## 3.1 Magnetic ground-state and equilibrium structural parameters

We have investigated several possible magnetic orders of the moments on the Cr atoms, either ferromagnetic (FM), antiferromagnetic (AFM), or with no magnetic moments (NM). As reported earlier, in the case of *GGA*, the ground state might correspond to either NM [30] or AFM [8], while for *GGA+U* the antiferromagnetic spin distribution of Cr atoms along the c-axis turns out to be the most stable solution [30]. In particular, using the functional (present study) we also found small amounts of localized Cr magnetic moments with an AFM spin ordering inside the in-plane Cr-C networks (i.e., $Cr_1^{(\uparrow)}$=+0.012 $\mu_B$, $Cr_2^{(\downarrow)}$=-0.008 $\mu_B$, $Cr_3^{(\downarrow)}$=-0.012 $\mu_B$, and $Cr_4^{(\uparrow)}$=+0.008 $\mu_B$). However, when using the +*U* corrected functional ($GGA^{WC}$+U), the ground state magnetism turns out to be rather different, having an alternate FM spin distribution for the two Cr-C networks (Fig. 1). The computed magnetic moments for the Cr atoms that belong to the Cr-C network located at nearly half of the *c*-axis are $Cr_2^{(\uparrow)}$=+0.011 $\mu_B$ and $Cr_4^{(\uparrow)}$=+0.007 $\mu_B$, whereas for those at the bottom/top of the hexagonal unit cell amount to $Cr_1^{(\downarrow)}$=-0.007 $\mu_B$ and $Cr_3^{(\downarrow)}$=-0.011 $\mu_B$. As shown in Fig. 1, the +*U* corrected functional allows for a Ge-mediated super-exchange magnetic coupling [31] between Cr atoms belonging to different Cr-C networks (Fig. 2).

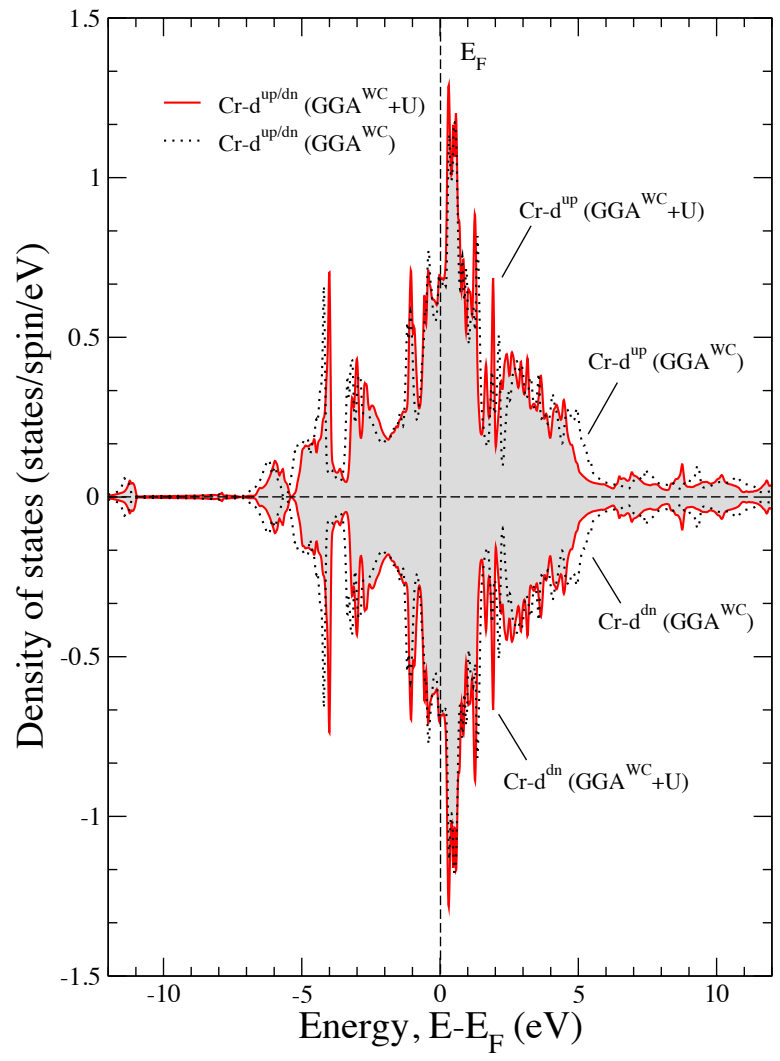

**Figure 3:** Density of states of AFM $Cr_2GeC$ for spin-up (upper part) and spin-down (lower part) Cr-*d* electrons.

This inter-layer interaction arises from the mixing of the Cr 3*d* and the Ge 4*p* states, which act along the 103° angle bend Cr-Ge-Cr three-point line. The computed unequal values for the Cr magnetic moments ($Cr_1 \neq Cr_2$ and $Cr_3 \neq Cr_4$) are ascribed to the small amount of $Ge_1 \rightarrow Cr_4$ and $Ge_2 \rightarrow Cr_1$ charge-transfer that is at the base of the super-exchange coupling mechanism [31]. This generates a very small spin polarization of the Ge atoms that sustains an antiferromagnetic spin coupling with the Cr atoms that are non directly involved in the charge-transfer mechanism ($Ge_1^{(\uparrow)}$-$Cr_3^{(\downarrow)}$ and $Ge_2^{(\downarrow)}$-$Cr_2^{(\uparrow)}$). Also, when introducing the on-site Coulombic interaction the three C atoms, which constitute the first coordination shell of Cr ions, become slightly spin-polarized, thus stabilizing a FM in-plane Cr spin distribution ($Cr_{1,3}^{(\downarrow)}$-$3\times C_2^{(\uparrow)}$ and $Cr_{2,4}^{(\uparrow)}$-$3\times C_1^{(\downarrow)}$). It is worth noting that the total magnetic moment of the unit cell is still zero, as to indicate an almost perfect resulting AFM spin alignment inside the whole hexagonal lattice. Despite the tiny magnitude of Cr magnetism computed within the GGA$^{WC}$+U scheme, the achieved results are clearly showing that $Cr_2GeC$ has a considerably more complicated magnetic structure than believed earlier. Relativistic corrections in the electronic structure calculation have also been included in a second-variational procedure using scalar-relativistic wavefunctions [32]. Applying the spin-orbit coupling within the atomic spheres along the [0 0 1] magnetization direction, leads to enhanced Cr magnetic moments while keeping exactly the same non-relativistic ground-state spin pattern.

The importance of the super-exchange coupling has been further underlined by GGA$^{WC}$+U calculations performed on an iso-geometrical Ge-hollow unit cell. In this case, an AFM material is stabilized with exactly the same in-plane AFM spin arrangement of Cr atoms found within the GGA$^{WC}$ method. Therefore, only an explicit description of the strongly correlated nature of the Cr 3*d* electrons is able to catch the super-exchange interaction, that maintain an in-plane FM Cr spin ordering. Preliminary experimental results seem to confirm this important finding [33], although the presence of small Cr magnetic moments might lead to a rather weak magnetic energy and therefore to a low Néel temperature.

Attention should also be payed to the nature of the interleaved non-magnetic *A*-atoms, that might play a crucial role in determining the overall magnetic properties of this *MAX*-phase. As a rule of thumb, the smaller the spin polarizability of the bridging *A*-atom is, the weaker the inter-layer super-exchange coupling will be [34]. Hence, smaller atoms having tightly bound valence electrons will then tend to weaken the super-exchange coupling, enhancing the FM behavior of the $Cr_2GeC$ crystal phase. Formulated differently, this will translate into attempting to tune the Cr-*A* bonding type as to reduce its covalent character, thus rising up the observable Néel temperature.

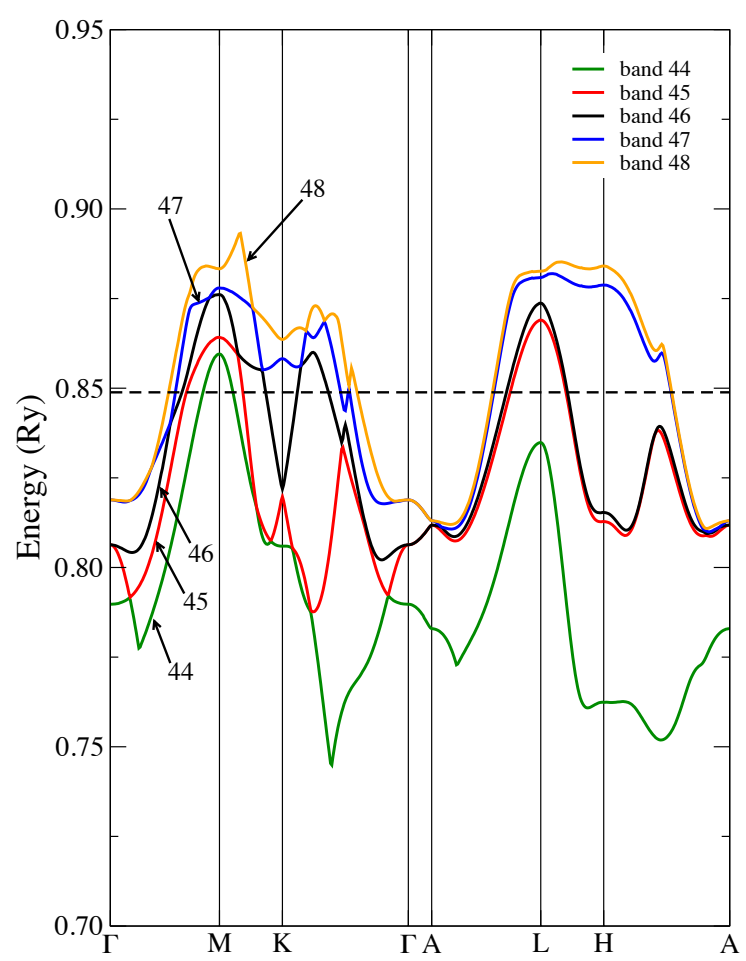

**Figure 4:** Energy band structure of $Cr_2GeC$ for the spin-up configuration along high symmetry directions shown in Fig. 5. Only the electronic bands that are crossing the Fermi level are shown.

In table 2, we report the volume, lattice parameters and bulk modulus within different correlation corrected *xc*-functionals. After including the $+U$ correction, the equilibrium volume considerably increases giving a good agreement with the experimental data of Phatak *et al.*[14], although the bulk modulus gets significantly lower. However, since the is correctly treating the correlated nature of the Cr *d*-electrons, we believe that most likely this is the right value for the bulk modulus.

**Table 2:** Optimized cell parameters for the ground-state AFM spin configuration.

| Property | $GGA^{WC}$ | $GGA^{WC}+U$ | $GGA^{PBE}+U^{a}$ | Exp.[b] | Exp.[c] |
|---|---|---|---|---|---|
| $V$ (Å$^3$) | 86.24 | 92.71 | 91.21 | 91.10 | 92.82 |
| $a$ (Å) | 2.899 | 2.981 | 2.97 | 2.950 | 2.958 |
| $c$ (Å) | 11.875 | 12.044 | 12.16 | 12.086 | 12.249 |
| $c/a$ | 4.097 | 4.040 | 4.094 | 4.097 | 4.141 |
| $B_0$ (GPa) | 254.4 | 147.6 | 150 | 182 | 169 |

[a] Using $U = 1.95$ eV and $J = 0.95$ eV [30].
[b] From [13].
[c] From [14].

### 3.2 Electronic density of states and band structure

The DOS at the Fermi level is dominated by the Cr transition metal. Fig. 3 illustrates the calculated Cr-*d* total DOS for both $GGA^{WC}$ and $GGA^{WC}$+U methods. A bandwidth reduction of about 0.75 eV is found when including the on-site Coulombic interaction. The most distinct feature we observe is the shrinking of both valence and conduction bandwidths. The bottom of the valence band moves to higher energy by 0.25 eV and the top of the conduction band is reduced towards lower energies by 0.50 eV. The Cr *d*-electrons strongly hybridize with the Ge and C *p*-states and their relative energetic position determines the degree of hybridization and the width of the valence band. The

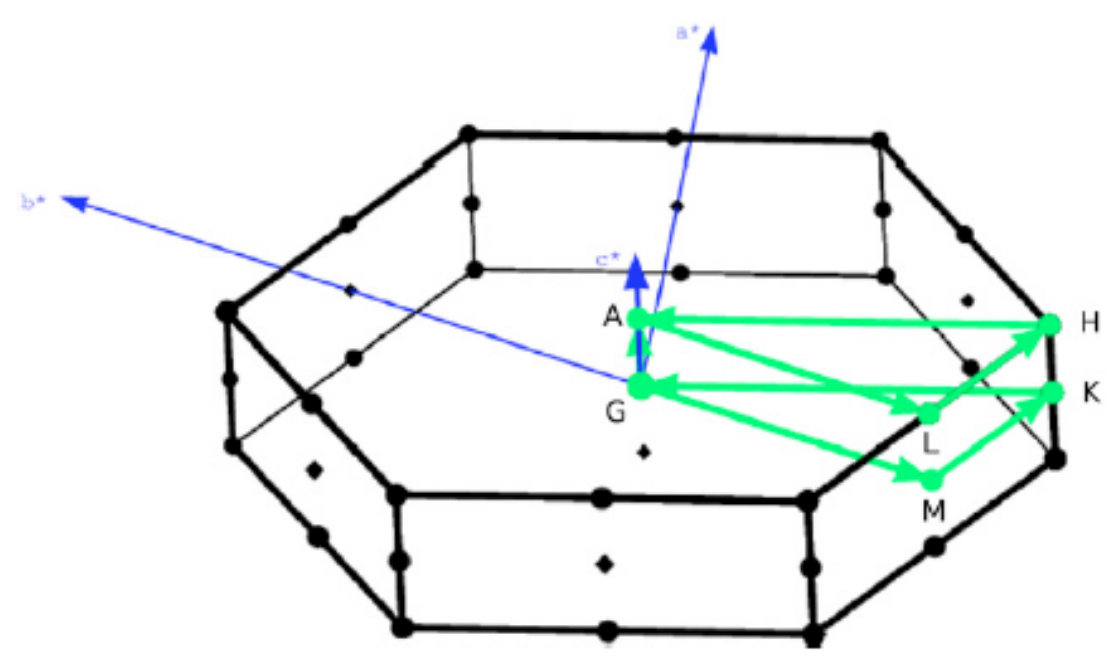

**Figure 5:** Primitive Brillouin zone of the hexagonal unit cell with the used symmetry points: Γ (0,0,0), *M* (1/2,0,0), *K* (2/3,1/3,0), *A* (0,0,1/2), *L* (1/2,0,1/2) and *H* (2/3,1/3,1/2). Reciprocal lattice vectors are also shown as a* ($k_x$), b* ($k_y$), and c* ($k_z$).

Hubbard *U* correction has therefore a noticeable influence on the hybridization between localized and itinerant states. The top of the valence band and the bottom of the conduction band develop more C 2*p* and Ge 3*p* characters when including *U* and the Cr states hybridizes accordingly giving rise to smaller bandwidths. A similar behavior of Ge states was found on the $V_2GeC$ phase [35].

Since $Cr_2GeC$ is a metallic system, the DOS at the Fermi level is a key quantity for stability purposes. The $Cr_2GeC$ crystal has its $E_F$ positioned exactly at a local minimum in the DOS, thus suggesting a higher level of intrinsic stability. As a matter of fact, local minimum at E is a good indicator of large structural stability as it represents a barrier for electrons below the Fermi-level to move into the unoccupied empty states. The $GG^{-WC}$+U method slightly increases the number of states at $E_F$ (3.95 states/eV) with respect to calculations (3.85 states/eV), while keeping the same topological DOS. As such, band renormalization does not provide any remarkable effects concerning the total amount of electronic band filling of the occupied states.

Figure 4 shows the $GGA^{WC}$+U calculated electronic band structure of $Cr_2GeC$. The dominant contribution to the electronic density of states at $E_F$ derives from metallic bonding of the Cr *d*-electron orbitals in the Cr-Ge-C network. Several bands are formed that cross E, both electron- and hole-like, thus resulting in a multiband system dominated by Cr *d*-character. At the Fermi level, the bands that are crossing the Fermi energy are 44, 45, 46, 47 and 48, and their numbering follows exactly that of Fig. 4. Band 44 has the same $d_{xz}+d_{yz}$ orbital character contribution from all the Cr atoms ($Cr_1 \rightarrow Cr_4$), while in band 45 there is an important $dz^2$ weight from $Cr_1$ and $Cr_2$ and a $d_{x2-y2}+d_{xy}$ contribution from $Cr_3$ and $Cr_4$. For band 46 all the Cr atoms contribute with the same amount of both $dz^2$ and $d_{x2-y2}+d_{xy}$ characters. The very similar bands 47 and 48 are mainly of $d_{x2-y2}+d_{xy}$ character from all the Cr atoms. Figure 4 also shows the hole- and electron-like character of the crossing bands. Specifically, hole-like features are positioned at symmetry point *M*, midway the K-Γ symmetry line, and at *L* and *H*. On the contrary, an electron-like pocket can be seen at symmetry points Γ, *A* and *K*.

From band structure analysis, important information about electronic transport properties can be obtained. The Cr-containing *MAX* carbides are expected to show the highest resistivity along the series $Ti_2GeC \rightarrow V_2GeC \rightarrow Cr_2GeC$. As a matter of fact, the reported resistivity for Ti- and V-based *MAX* phases are in the range 15-30 μΩcm compared to 53-67 μΩcm for the $Cr_2GeC$ material [5]. This is generally due to the reduced carrier mobilities in Cr-based systems, where strongly localized Cr *d*-states are present near the Fermi level [36]. In this regard, it is worth noting that the overall electronic band structure of $Cr_2GeC$ is rather anisotropic, with bands crossing the E only along the symmetry lines of the basal *xy*-plane. Nevertheless, such an intrinsic band structure anisotropy is similar to that of typical hexagonal-close-packed materials [37], and therefore cannot be used to quantitatively explain the peculiar transport properties of Cr-containing *MAX* phases. In this respect, scattering processes and charge carrier-phonon coupling should be taken into account [38].

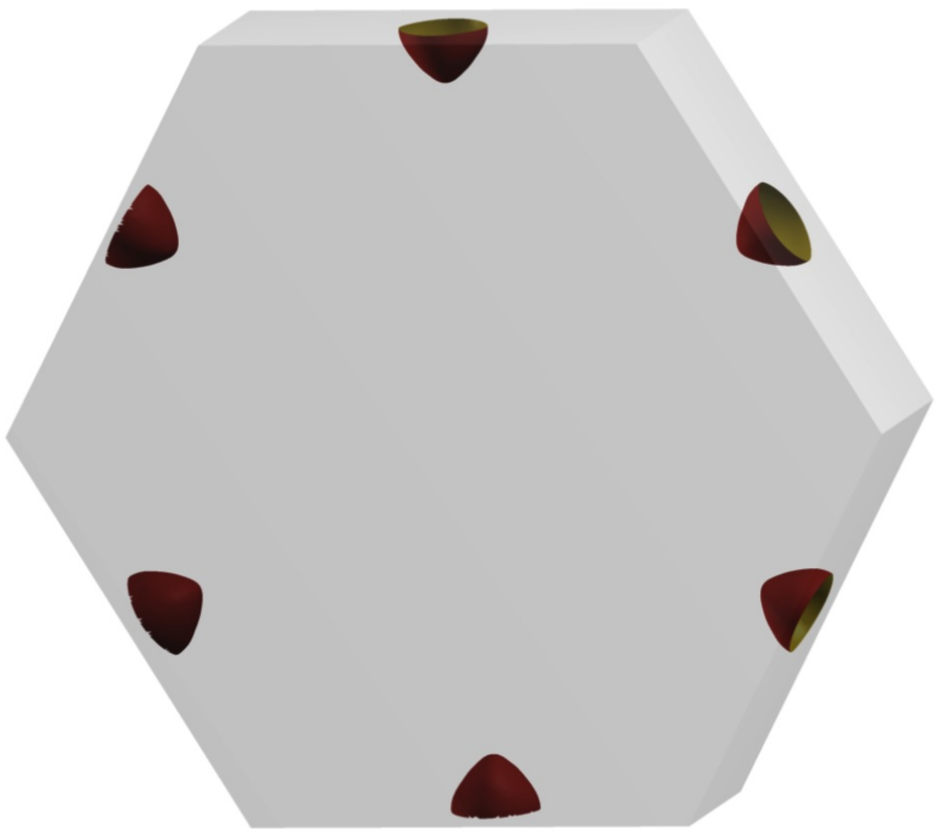

**Figure 6:** Constant energy surface for band 44 (spin-up) viewed along the $k_x$ and $k_y$ plane of the hexagonal Brillouin zone.

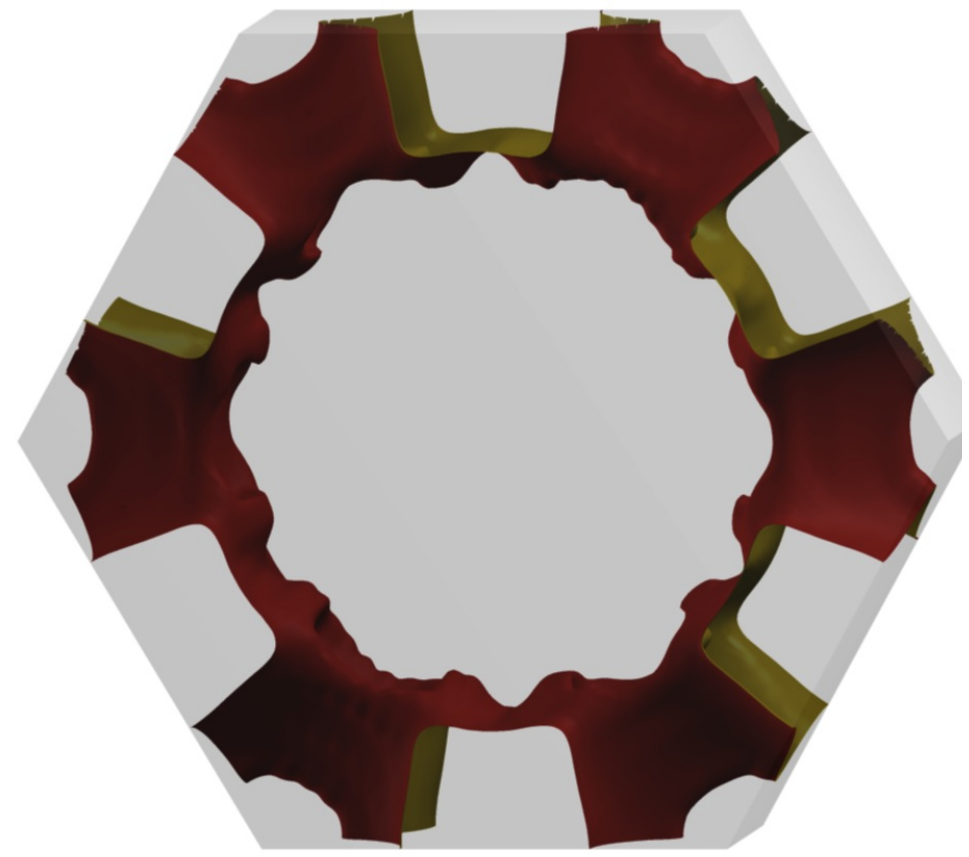

**Figure 8:** Fermi surface for band 46 (spin-up).

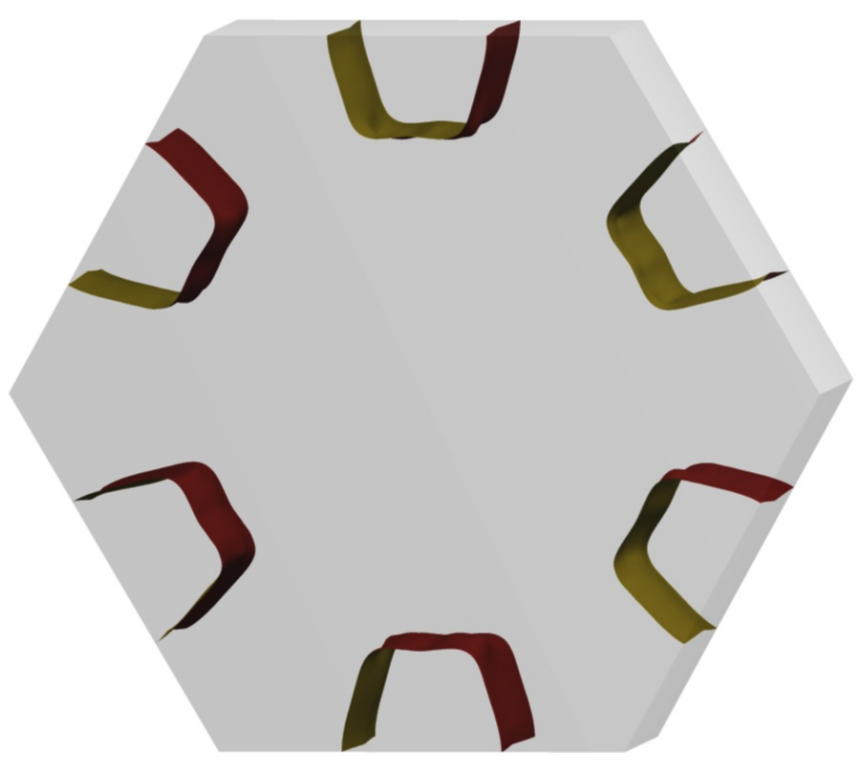

**Figure 7:** Fermi surface for band 45 (spin-up).

### 3.3 Fermi surfaces

In metals, the energy states that participate in determining most properties of a material lie in close proximity to the Fermi energy, that is, the level below which available energy states are filled. The Fermi surface thus separates the unfilled orbitals from the filled ones and represents a surface of constant energy ($E=E_F$) in **k**-space. The electrical properties of metals are defined by the shape and size of the Fermi surface, as the current is due to changes in the occupancy of states, near the Fermi surface. Therefore, by observing the *fermiology* of the computed Fermi surfaces one can help addressing, from a qualitative point of view, the predominant role of each crossing band along either the *z* component or inside the *xy* plane.

Since the velocities of the electrons are perpendicular to the Fermi surface, then bands 45, 47 and 48 have large components within the basal plane, while bands 44 and 46 play an equally important role along the three $k_x$, $k_y$, and $k_z$ axes. From band 44 a localized hole-like FS pocket emerges at the M-point (Fig. 6), defined as the **k**-vector (1/2,0,0) in the BZ of Fig. 5. The next band (45) has two hole-like band features, one centered at the symmetry point *M* and the other at *L* (Fig. 7). Band 46 (Fig. 8) has a mixture of characters, with hole-like features at *M* and *L* and an electron-like component at *K*.

The Fermi surfaces of bands 47 and 48 (Figs. 9 and 10, respectively) have both very similar hole-like character along the point symmetries Γ-M, K-Γ, A-L and A-H.

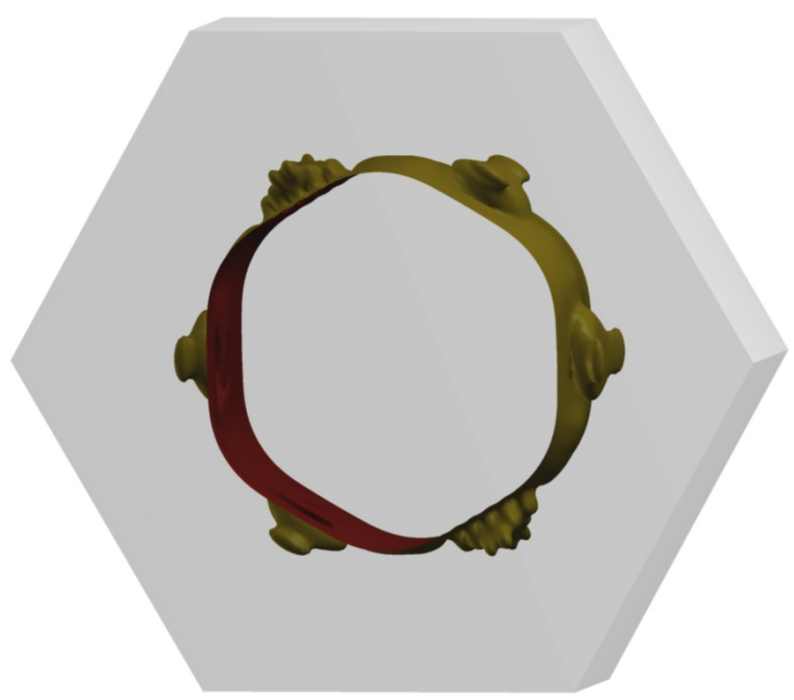

**Figure 9:** Fermi surface for band 47 (spin-up).

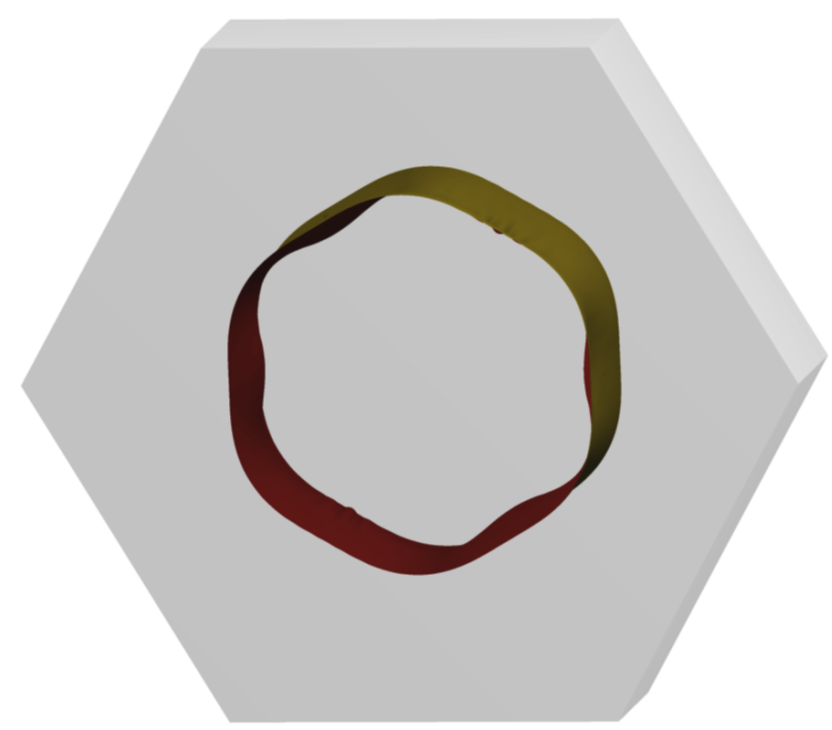

**Figure 10:** Fermi surface for band 48 (spin-up).

As with other transition metal-based materials, the introduced electronic correlation leads to a certain degree of electron renormalization of the band structure, which, however, does not change the main topology of the $Cr_2GeC$ Fermi surfaces.

## 4. Discussion

It has been shown that an appropriate treatment of correlation effects in the $Cr_2GeC$ *MAX*-phase leads to the discovery of a different magnetic spin pattern, where a super-exchange interaction operates through the non-magnetic Ge ions. The buckled Cr-C networks that are propagating parallel to the *xy*-plane of the crystal are thus showing a FM intra-layer spin distribution and an AFM inter-layer spin ordering. The FM layers are lined up in a perfect anti-ferromagnetic pattern giving a vanishing total magnetic moment inside the unit cell. This finding enables the possibility of tuning the exchange coupling between ferromagnetically ordered Cr-C layers so as to achieve stable magnetic *MAX* phases for electronic and spintronic applications. For instance, the inter-layer coupling could be varied by changing the inter-layer thickness [39], interface quality [40] or even by alloying with another M-element [41]. Further studies are being pursued in order to experimentally confirm such a magnetic pattern via a detailed XMCD analysis [33].
From the calculated Fermi surfaces we have seen that there is only one electron-like band (band 46) that shows a velocity component along the k axis. The other 4 Fermi surfaces are hole-like with large velocity contributions confined within the basal xy-plane. Therefore, $Cr_2GeC$ appears to be a material characterized by a clear *carrier-type anisotropy*, being the positively hole charge carriers responsible for transport properties within the basal *xy*-plane, and the negatively electron charge carriers along the vertical *z*-axis. This qualitatively explains the reason why the experimentally determined Seebeck coefficients [33] along the [001] plane (i.e., the in-plane component) are generally larger than those along the [103] plane (i.e., the out-of-plane direction). Large and positive Seebeck coefficients indicate that $Cr_2GeC$ behaves as a *p*-type material along the in-plane direction, having predominantly positive mobile charges (holes). On the contrary, the much lower magnitude of measured Seebeck coefficients along the [103] plane points to an increased negative carrier concentration along the vertical direction of the hexagonal $Cr_2GeC$ crystal. This may also be true in other Cr-based materials such as $Cr_2AlC$. As shown in this work, determining the super-exchange coupling will serve as an important method to predict which *MAX* phases are good candidates with stable magnetic features.

## 5. Conclusions

An *ad hoc* effective Hubbard *U* value has been computed for various exchange correlation functionals by using the constrained DFT formalism. We have shown that $Cr_2GeC$ is a weak AFM material with a rather anisotropic electronic band structure. Most important, by properly accounting for Cr correlation effects we discovered the presence of a super-exchange coupling between different in-plane Cr-C networks of the $Cr_2GeC$ unit cell. Therefore, the magnetic nature of the studied *MAX* phase is AFM (as proposed earlier) but with a substantially different electro-structural origin. The interleaved Ge-atoms stabilize the ferromagnetically ordered Cr layers that are exchange coupled together. If this kind of inter-layer coupling can be tailored, then very attractive layered magnetic materials can be proposed with a potential use for many electronics and spintronics applications.
Equilibrium structural parameters were also computed within the $GGA^{WC}$+U and found to be in good agreements with the experimental data of Phatak *et al.*[14]. The topology of Fermi surfaces was studied to address the electric transport properties of the metallic $Cr_2GeC$ material. The achieved results indicate that this Cr-containing *MAX* phase has a relevant asymmetrical carrier-type structure, where hole carriers dominate within the basal plane and electrons only contribute to carrier mobility along the *z*-axis. We thank the Swedish Research Council (VR) for financial support.

## References

[1] Anisimov V I and Gunnarsson O 1991 Density-functional calculation of effective Coulomb interactions in metals *Phys. Rev. B* 45 7570-7574.
[2] Barsoum M W 2000 The M(N+1)AX(N) phases: A new class of solids; Thermodynamically stable nanolaminates *Prog. Solid State Chem.* 28 201-281.
[3] Wolf S A, Awschalom D D, Buhrman R A, Daughton J M, von Molnar S, Roukes M L, Chtchelkanova A Y, and Treger D M 2001 Spintronics: A spin-based electronics vision for the future *Science* 294 1488-1495.
[4] Barsoum M W and Radovic M 2011 Elastic and Mechanical Properties of the MAX Phases *Annual Review of Materials Research* 41 195-227.
[5] Eklund P, Beckers M, Jansson U, Hoegberg H, and Hultman L 2010 The M(n+1)AX(n) phases: Materials science and thin-film processing *Thin Solid Films* 518 1851-1878.
[6] Wang J and Zhou Y 2009 Recent Progress in Theoretical Prediction, Preparation, and Characterization of Layered Ternary Transition-Metal Carbides *Annual Review of Materials Research* 39 415-443.
[7] Bouhemadou A 2009 Calculated structural, electronic and elastic properties of MGeC (M=Ti, V, Cr, Zr, Nb, Mo, Hf, Ta and W) *Appl. Phys. A* 96 959-967.
[8] Zhou W, Liu L, and Wu P 2009 First-principles study of structural, thermodynamic, elastic, and magnetic properties of Cr2GeC under pressure and temperature *J. Appl. Phys.* 106 033501-033508.
[9] Scabarozi T H, Amini S, Leaffer O, Ganguly A, Gupta S, Tambussi W, Clipper S, Spanier J E, Barsoum M W, Hettinger J D and Lofland S E 2009 Thermal expansion of select M > n+1 > AX > n (M=early transition metal, A=A group element, X=C or N) phases measured by high temperature x-ray diffraction and dilatometry *J. Appl. Phys.* 105 013543-013551.
[10] Barsoum M, Scabarozi T H, Amini S, Hettinger J D and Lofland S E 2011 Electrical and Thermal Properties of Cr2GeC *J. Am. Ceram. Soc.* 94 4123-4126.
[11] Drulis M K, Drulis H, Hackemer A E, Leaffer O, Spanier J, Amini S, Barsoum M W, Guilbert T and El-Raghy T 2008 On the heat capacities of TaAlC, TiSC, and Cr2GeC *J. Appl. Phys.* 104 023526-023533.
[12] Eklund P, Bugnet M, Mauchamp V, Dubois S, Tromas C, Jensen J, Piraux L, Gence L, Jaouen M and Cabioch T 2011 Epitaxial growth and electrical transport properties of Cr2GeC thin films *Phys. Rev. B* 84 075424-075433.
[13] Manoun B, Amini S, Gupta S, Saxena S K and Barsoum M W 2007 On the compression behavior of Cr2GeC and VGeC up to quasi-hydrostatic pressures of 50 GPa *Phys. Condens. Matter.* 19 456218-456225.
[14] Phatak N A, Kulkarni S R, Drozd V, Saxena S K, Deng L, Fei Y, Hu J and Ahuja R 2088 Synthesis and compressive behavior of Cr2GeC up to 48 GPa *J. Alloy. Comp.* 463 220-225.
[15] Momma K and Izumi F 2008 VESTA: a three-dimensional visualization system for electronic and structural analysis *J. Appl. Crystallogr.* 41 653-658.
[16] Blaha P, Schwarz K, Madsen G K H, Kvasnicka D and Luitz J 2001 An Augmented Plane Wave + Local Orbitals Program for for Calculating Crystal Properties (Karlheinz Schwarz, Techn. Universität Wien, Austria), ISBN 3-9501031-1-2.
[17] Hohenberg P and Kohn W 1964 Inhomogeneous Electron Gas *Phys. Rev.* 136 B864-B871.
[18] Kohn W and Sham L J 1965 Self-Consistent Equations Including Exchange and Correlation Effects *Phys. Rev.* 140 A1133-A1138.
[19] Wu Z and Cohen R 2006 More accurate generalized gradient approximation for solids *Phys. Rev. B* 73 235116-235122.
[20] Tran F, Laskowski R, Blaha P and Schwarz K 2007 Performance on molecules, surfaces, and solids of the Wu-Cohen GGA exchange-correlation energy functional *Phys. Rev. B* 75 115131-115145.
[21] Blöchl P E, Jepsen O and Andersen O K 1994 Improved tetrahedron method for Brillouin-zone integrations *Phys. Rev. B* 49 16223-16233.
[22] Monkhorst H J and Pack J D 1976 Special points for Brillouin-zone integrations *Phys. Rev. B* 13 5188-5192.
[23] Kokalj A 2003 Computer graphics and graphical user interfaces as tools in simulations of matter at the atomic scale *Comp. Mater. Sci.* 28 155-168.

[24] Hubbard J 1963 Electron correlations in narrow energy bands *Proc. R. Soc. London* Ser. A 276 238.

[25] Anisimov V I, Solovyev I V, Korotin M A, Czyżyk M T and Sawatzky G A 1993 Density-functional theory and NiO photoemission spectra *Phys. Rev. B* 48 16929-16934.

[26] Dudarev S L, Botton G A, Savrasov S Y, Humphreys C J and Sutton A P 1998 Electron-energy-loss spectra and the structural stability of nickel oxide:An LSDA+U study *Phys. Rev. B* 57 1505-1509.

[27] Czyżyk M T and Sawatzky G A 1994 Local-density functional and on-site correlations: The electronic structure of LaCuO and LaCuO *Phys. Rev. B* 49 14211-14228.

[28] Petukhov A G, Mazin I I, Chioncel L and Lichtenstein A I 2003 Correlated metals and the LDA+U method *Phys. Rev. B* 67 153106-153110.

[29] Şaşioğlu E, Friedrich C and Blügel S 2011 Effective Coulomb interaction in transition metals from constrained random-phase approximation *Phys. Rev. B* 83 121101(R)-121105(R).

[30] Ramzan M, Lebégue S and Ahuja R 2012 Electronic and mechanical properties of Cr2GeC with hybrid functional and correlation effects *Solid State Comm.* 152 1147-1149.

[31] Anderson P W 1950 Antiferromagnetism. Theory of Superexchange Interaction *Phys. Rev.* 79 350-356.

[32] Singh D 1994 Plane waves, pseudopotentials and the LAPW method, Kluwer Academic, Boston.

[33] Magnuson M, Mattesini M, Bugnet M, Mauchamp V, Cabioch T, Hultman L and Eklund P 2012 *in manuscript*.

[34] Sherman D M 1985 The electronic-structures of Fe-3+ coordination sites in iron-oxides-applications to spectra, bonding, and magnetism *Phys. Chem. Minerals* 12 161-175.

[35] Magnuson M, Wilhelmsson O, Mattesini M, Li S, Ahuja R, Eriksson O, Högberg H, Hultman L, and Jansson U 2008 Anisotropy in the electronic structure of VGeC investigated by soft x-ray emission spectroscopy and first-principles theory *Phys. Rev. B* 78 035117-035126.

[36] Hettinger J D, Lofland S E, Finkel P, Meehan T, Palma J, Harrel K, Gupta S, Ganguly A, El-Raghy T, and Barsoum M W 2005 Electrical transport, thermal transport, and elastic properties of MAlC (M=Ti, Cr, Nb, and V) *Phys. Rev. B* 72 115120-115126.

[37] Sanborn B A, Allen P B and Papaconstantopoulus D A 1989 Empirical electron-phonon coupling constants and anisotropic electrical resistivity in hcp metals *Phys. Rev. B* 40 6037-6044.

[38] Magnuson M, Mattesini M, Nong N V, Eklund P and Hultman L 2012 Electronic-structure origin of the anisotropic thermopower of nanolaminated TiSiC determined by polarized x-ray spectroscopy and Seebeck measurements *Phys. Rev. B* 85 195134-195142.

[39] Grünberg P, Schreiber R, Pang Y, Brodsky M B and Sowers H 1986 Layered Magnetic Structures: Evidence for Antiferromagnetic Coupling of Fe Layers across Cr Interlayers *Phys. Rev. Lett.* 57 2442-2445.

[40] Ikeda S, Hayakawa J, Ashizawa Y, Lee Y M, Miura K, Hasegawa H, Tsunoda M, Matsukura F and Ohno H 2008 Tunnel magnetoresistance of 604% at 300 K by suppression of Ta diffusion in CoFeB/MgO/CoFeB pseudo-spin-valves annealed at high temperature *Appl. Phys. Lett.* 93 082508-082511.

[41] Dahlqvist M, Alling B, Abrikosov I A and Rosen J 2011 Magnetic nanoscale laminates with tunable exchange coupling from first principles *Phys. Rev. B* 84 220403(R)-220408(R).